# REVERSIBLE FERROMAGNETIC SWITCHING

# IN ZnO:(Co, Mn) POWDERS

D. Rubi[*] and J. Fontcuberta

Institut de Ciència de Materials de Barcelona, Campus UAB, E-08193, Bellaterra, Spain

A. Calleja and Ll. Aragonés

Quality Chemicals, Fornal 35, Pol. Ind. Can Comelles Sud, Esparreguera, 08292, Spain

X.G. Capdevila and M. Segarra

Facultat de Química, Universitat de Barcelona, Martí i Franquès 1, 08028, Barcelona, Spain

We report here on the magnetic properties of ZnO:Mn and ZnO:Co doped nanoparticles. We have found that the ferromagnetism of ZnO:Mn can be switched on and off by consecutive low-temperature annealings in $O_2$ and $N_2$ respectively, while the opposite phenomenology was observed for ZnO:Co. These results suggest that different defects (presumably n-type for ZnO:Co and p-type for ZnO:Mn) are required to induce a ferromagnetic coupling in each case. We will argue that ferromagnetism is likely to be restricted to a very thin, nanometric layer, at the grain surface. These findings reveal and give insight into the dramatic relevance of surface effects for the occurrence of ferromagnetism in ZnO doped oxides.

[*] Currently at Materials Science Center, University of Groningen, Groningen 9747AG, The Netherlands.
E-mail: d.rubi@rug.nl

## 1. Introduction

The development of diluted magnetic semiconductors (DMS) with Curie temperatures above room temperature could eventually lead to a new generation of spintronic devices with exciting novel functionalities. Early theoretical claims [1,2] attributed this potential to TM:ZnO (where TM is a transition metal), stimulating intensive studies on the magnetic properties of ZnO doped with different magnetic cations. However, the results obtained up to now have been very controversial. Indeed, the observation of room-temperature ferromagnetism has been attributed either to an intrinsic [3,4] or extrinsic [5,6] property of TM:ZnO, while other reports claim for a genuine paramagnetic behaviour [7,8]. This indicates that an intense effort on the understanding of the magnetic interaction of these materials remains to be done.

In agreement with some theoretical predictions [2], recent experiments suggest that n- and p-type defects mediate the ferromagnetism of Co:ZnO and Mn:ZnO compounds [4,9], respectively, raising the possibility of tailoring the ferromagnetic interaction of these materials by modifying their defect structure. For instance, it has been shown that the ferromagnetism of both epitaxial [4,10] and nanoparticles thin films [11,4] of Co:ZnO can be reinforced by exposing them to Zn vapour, which was recently shown to generate zinc interstitials ($Zn_i$) n-type defects [12]. On the other hand, fast annealings of Mn:ZnO nanocrystals capped with N-rich molecules [4] have been shown to enhance their ferromagnetism. It has been claimed that this effect results from the introduction of p-type $N_O$ defects. The aforementioned correlation between ferromagnetism and point-defects strongly supports the bound magnetic polaron (BMP) model for ferromagnetism in diluted magnetic semiconductors [13]. According to this model, the magnetic exchange among magnetic impurities is mediated by carriers in a split-spin impurity band derived from extended donor orbitals. The opposite polarity of the carriers necessary to mediate the ferromagnetic coupling in Co:ZnO and Mn:ZnO has been related to the different positions of the magnetic impurities and point-defects energy levels in the ZnO band gap, which allows an efficient magnetic impurity-defect hybridization only in the cases of n-type Co:ZnO and p-type Mn:ZnO [9].

Once settled the key role played by point-defects on the stabilization of the ferromagnetism of ZnO-based diluted magnetic semiconductors, it is evident that the achievement of intrinsic ferromagnetism in these materials is strongly linked to the ability to control their point-defects structure. Here we will show that the ferromagnetism of Co:ZnO and Mn:ZnO nanopowders can be successively switched on and off by low-temperature annealings under suitable atmospheres. The obtained results demonstrate both the correlation between defects and ferromagnetism *and* the distinct polar behaviour of Co:ZnO and Mn:ZnO. We will argue that the modification of the electronic structure of the ZnO surface, either by point defects or gas chemisorption, controls the sample magnetization.

**2. Experimental**

Several series of nanopowders, with nominal stochiometries $Co_{0.1}Zn_{0.9}O$ and $Mn_{0.1}Zn_{0.9}O$, were prepared by means of the acrilamide polymerization method (see Ref. [14] for a detailed description of the preparation method). The obtained xerogels after self-combustion (labelled ZC1 and ZM1 for Co and Mn:ZnO, respectively) were grinded and fired in air for 12 hours in a muffle furnace at temperatures $T_S$ of 300ºC, 400ºC, 500ºC and 600ºC. Here we will present representative results corresponding to the series labelled as ZC1_$T_S$ and ZM1_$T_S$, where $T_S$ is the final firing temperature. Structural characterization was performed by X-ray diffraction (Siemens D-5000, using CuK$_{\alpha1,\alpha2}$ radiation) and high resolution transmission electronic microscopy (FEG TEM Jeol 2019F). The particle size was estimated from the broadening of the XRD peaks, obtaining crystallite sizes of ~ 40nm. The cationic composition was checked by energy dispersive X-ray spectroscopy (EDS), finding in all cases Mn/Zn and Co/Zn ratios consistent with the nominal stoichiometry. Magnetic measurements were performed by means of a Superconducting Quantum Interference Device from Quantum Design.

### 3. Results and Discussion

Figure 1(a) shows the XRD patterns corresponding to ZM1 and ZC1 samples fired at $T_S$ between 300ºC and 600ºC. The spectra are dominated by reflections corresponding to the hexagonal wurtzite structure of ZnO (S.G. *P6$_3$ mc*); however, the XRD patterns reveal that increasing the calcination temperature ($T_S$) favours the emergence of additional peaks (marked with *) that reflect the segregation of the oxide $ZnMnO_3$ and the spinel $Zn_xCo_{3-x}O_4$, respectively. No additional reflections associated to other segregated phases were detected. The amount of segregated $Zn_xCo_{3-x}O_4$ was evaluated from the Rietveld refinements of the corresponding XRD patterns. In the case of $ZnMnO_3$, Rietveld refinements are not possible since its structure remains undetermined [15]. In consequence, the concentration of $ZnMnO_3$ as a function of $T_S$ has been simply estimated from the ratio between the ~30.2º peak of $ZnMnO_3$ and the ~31.7º peak of ZnO. Figure 1(b) clearly shows the progressive segregation of $Zn_xCo_{3-x}O_4$ and $ZnMnO_3$ when increasing $T_S$. The cell volume of the wurtzite phase of the ZM1 series was found to shrink from 48.11(1) Å$^3$ (ZM1_300 sample) to 47.80(1) Å$^3$ (ZM1_600 sample) when increasing $T_S$. Bearing in mind the $Mn^{2+}$ and $Zn^{2+}$ ionic radii (0.66 Å and 0.60Å, respectively), the cell contraction reflects the migration of the larger Mn cations out from the wurtzite structure to form the $ZnMnO_3$ impurity. In the case of Co-doped samples, we have found that the cell volume remains in all cases close to the bulk ZnO value (47.62Å$^3$). This is related to the similar ionic radii of $Co^{2+}$ and $Zn^{2+}$ (0.58 Å and 0.60 Å, respectively), which leads into a wurtzite unit cell rather insensitive to Co contents. Detailed high resolution TEM analysis [16] failed to detect other segregated phases than those observed by XRD. In particular, we should remark that the existence of segregated metallic Co or Mn was not appreciated in any case. Optical absorption and X-ray photoemission spectroscopy experiments, performed on ZM1_300 and ZC1_300 samples, suggest valence states of 2+ and tetrahedral coordination for both Co and Mn ions, indicating that they replace Zn ions in the wurtzite structure.

Figures 2(a) and 2(b) show the evolution of the room-temperature magnetization as a function of the applied field for both ZM1 and ZC1 series. The magnetic moments per atom were evaluated by considering the nominal concentration of magnetic ions. It is found that Mn:ZnO and Co:ZnO samples fired at 300ºC display a clear ferromagnetic response superimposed to a paramagnetic component. In contrast, samples treated at higher temperatures are paramagnetic-like. The saturation magnetizations of ZM1_300 and ZC1_300 samples are rather small (~0.01$\mu_B$/Co and ~0.002$\mu_B$/Mn, respectively). This observation, which is in agreement with many reported data for both bulk samples [17,18] and thin films [11,19], indicates that only a small fraction of the available magnetic ions are ferromagnetically coupled. The insets in Figures 2(a) and (b) show the temperature dependence of the inverse magnetic susceptibility ($\chi^{-1}$) for ZM1_500 and ZC1_500 samples; in agreement with the magnetization loops, data do not display any ferromagnetic signature. Following Ref. [7], the observed $\chi^{-1}(T)$ dependence can be well fitted by assuming that the susceptibility contains two contributions: (i) a Curie-Weiss term ($\chi_1 = C_1/T+\theta$, $\theta<0$) and (ii) a Curie term ($\chi_2 = C_2/T$). The first terms takes into account the antiferromagnetic exchange between near-neighbour magnetic ions (TM-O-TM bonds), while the second one describes the behaviour of isolated magnetic atoms. Solid lines through the data points in the insets of Figures 2(a) and 2(b) are the results of the fits; it can be appreciated that the model accurately reproduces the experimental data. For paramagnetic samples of the ZM1 series, the extrapolated temperatures $\theta$ were in the range ~ -120/-190 K, while for ZC1 series the corresponding values were in the range ~ -25/-50 K. The relative fractions of paramagnetic ions $C_2/(C_1+C_2)$ were found to be ~ 40-50% and ~ 25-35% for ZM1 and ZC1 series, respectively. The obtained $\theta$ values for Co-doped samples are larger than those obtained for Mn-doped samples, indicating a stronger antiferromagnetic coupling in the case of Mn-O-Mn bonds.

At this point, it is necessary to discuss about the origin of the ferromagnetic behaviour observed in the samples processed at low temperature (300ºC), paying special attention to the possible presence of extrinsic mechanisms. We first notice that the observed impurities ($Zn_xCo_{3-x}O_4$ and $ZnMnO_3$) can not account for a room-temperature ferromagnetic behaviour: the spinel $Zn_xCo_{3-x}O_4$ is ferrimagnetic at low temperatures ($T_C<50K$) [20], while $ZnMnO_3$ has been described as a spin glass with a blocking temperature of ~15K [21]. Moreover, Fig. 1b and Fig. 2 clearly show that the evolution with $T_S$ of both the fraction of segregated secondary phases and the sample magnetization follow opposite trends, thus strongly denying the observed secondary phases as responsible for the measured ferromagnetism. We recall that the usual origin of extrinsic ferromagnetism in Co:ZnO is metallic Co precipitates [5], which have not been observed in our samples neither by XRD nor HRTEM. Similarly, extrinsic ferromagnetism in Mn:ZnO has been proposed to arise from a metastable $(ZnMn)_2O_3$ phase [6] or from interface effects between Zn-rich $Mn_2O_3$ and $MnO_2$ interfaces [22]; none of these mechanisms appear to be dominant in our samples as we have failed to detect the presence of $Mn_2O_3$ or $MnO_2$ impurities. Therefore, we conclude that the observed ferromagnetism in our ZM1_300 and ZC1_300 samples is a genuine property of these oxides. This conclusion is also supported by the observation (see below) of reversible switching of ferromagnetism in these samples.

Interestingly enough, we have found that the as-grown ferromagnetism of ZM1_300 and ZC1_300 samples –which have been stored at room temperature for a period of ~ 2 months- vanishes with time, leading to a fully paramagnetic response [23]. We have not been able to observe any significant change in the high field slope of the hysteresis loops of ZM1_300 and ZC1_300 samples upon aging, indicating that the relative variation of the number of paramagnetic ions is rather small and any modification in their magnetic response remains below the experimental error. The magnetic aging was not accompanied by any detectable structural or chemical modification, suggesting that it may be related to variations of the samples defect structure. The time scale of the observed aging is consistent with reported photoluminescence (PL) experiments on pure ZnO samples, where it was shown that a spontaneous and continuous quenching of the PL green band (~2.5eV) – usually ascribed to the presence of point-defects [24]- takes place even one year after the sample fabrication [25]. On the other hand, it has been observed that the green PL band of pure ZnO can be modified in a faster a more controlled way by annealing under different atmospheres such as $O_2$ or $N_2$ [25,26]. We have therefore annealed the "magnetically aged" ZM1_300 and ZC1_300 samples, at 300ºC, under different atmospheres: air, $O_2$, $N_2$, or $H_2$-Ar (5%). Interestingly enough, we have found that the ferromagnetism of ZC1_300 sample can be regenerated *only* under $N_2$ atmosphere, as can be observed in Figure 3(a), while the ferromagnetism of ZM1_300 sample can be switched on *only* under $O_2$ annealing (Figure 3(b)). Figure 3(a) shows that the ferromagnetic component of ZC1_300 sample monotonically increases with the $N_2$ time annealing. After a 40hours treatment, the saturation magnetization raises from ~0 to ~$0.0018\mu_B$/Co. In the case of ZM1_300 sample, the enhancement of the ferromagnetism is not monotonic with the $O_2$ time annealing but displays a maximum for a 12h treatment. In this case, the saturation magnetization is of about ~$0.0026\mu_B$/Mn. Annealed samples were carefully analysed by XRD and HRTEM, finding in all cases that their structural and chemical properties remain unmodified. This supports the idea that only the defect structure of the samples is varied upon annealing. The regenerated ferromagnetism can be quenched again by performing further treatments under the "opposite" atmospheres: $N_2$ for Mn:ZnO and $O_2$ for Co:ZnO. This is illustrated in Figure 4(a), where it can be seen that successive 300ºC treatments under $O_2$ (12h) and $N_2$ (20h) atmospheres can switch on and off the ferromagnetism of Mn:ZnO, while the inverse phenomenology is found for Co:ZnO (Figure 4(b)).

The observed "polar" behaviour of Mn:ZnO and Co:ZnO (see Figure 4(c)) strongly reminds the suggested link between the occurrence of ferromagnetism in Co:ZnO and Mn:ZnO with the existence of n- and p-type defects, respectively [4,9]. Within this framework, it can be inferred that the n- and p-type character of our samples are improved by $N_2$ and $O_2$ annealing, respectively. In the first case, it can be argued that the inert character of the $N_2$ atmosphere may lead to the generation of donor oxygen vacancies ($V_O$). It follows that an analogous effect would be expected after similar annealing under other inert or reducing atmospheres such as Ar or $H_2$-Ar (5%). However, we have not been able to induce ferromagnetism with other gases than $N_2$, thus suggesting that $N_2$ should play an active role on the formation of n-type defects like, for example, the double donor $(N_2)_O$ [27]. In the case of $O_2$ annealing, the most plausible effect is the cancellation of native oxygen vacancies, which may reduce the n-type character of the material and eventually allow some previously compensated native acceptors to hybridize with Mn ions and mediate the ferromagnetic coupling.

Finally, we would like to address the key -but puzzling- observation of substantially reduced magnetization values with respect the expected transition metals moments. We recall that the observed values were ~$0.01\mu_B$/Co and ~$0.002\mu_B$/Mn for the as-grown ZC1_300 and ZM1_300 samples, while the expected values for $Co^{2+}$ and $Mn^{2+}$ high-spin ions are $3\mu_B$/Co and $5\mu_B$/Mn, respectively. We stress that rather marginal magnetization values have been frequently observed in reports dealing with polycrystalline samples [17, 18, 28, 29]. We first notice that a substantial magnetization reduction should be indeed expected due to the presence of antiferromagnetic coupled metal ions. In the case of ZC1_300 sample, for instance, if we assume as an upper-bound limit that Co spins displaying antiferromagnetic interactions (~75% of the total, as determined from the fittings of the $\chi^{-1}(T)$ data) do not contribute to the ferromagnetic signal, the estimated magnetization for this sample would be of about $0.04\mu_B$/Co; obviously this value is still considerably smaller than the expected one, suggesting that only a small fraction of the available Co ions interact ferromagnetically. The annealing process is performed at low temperature and thus the diffusivity of defects should be necessarily small. In the simplest picture, if we assume spherical ZnO particles with radius R~20nm and an overall magnetization $M_S$~0.04$\mu_B$/Co, a fully saturated ($M_S$= 3$\mu_B$/Co) thin ferromagnetic layer of about one ZnO unit cell is enough to account for the observed magnetization. For instance, if we use as an upper bound a diffusion coefficient of $O_2$ of about ~5 $10^{-17}$ cm$^2$/s (as determined at 900 ºC for pure ZnO polycrystals [30]), it is found that a processing time of 12h would give a profile with a characteristic depth (1/e decay) of about 10 nm. This rough estimate suggests that the defect structure modification upon 300ºC annealing can only take place in a very thin surface layer of ZnO (<<10nm), and it is just this layer which is modified upon annealing and originates the ferromagnetism. We should notice that the latter discussion has assumed that the defects created by annealing are point-defects located near the surface; however, we should stress that the existence of purely surface defects – arguably related to the chemisorption of $O_2$ or $N_2$ gas upon low temperature annealing- could also account for similar effects [31]. For instance, it was shown that holes can be induced in ZnO by $O_2$-chemisorption at low temperatures [31], which may explain the generation of ferromagnetism in our Mn:ZnO samples upon $O_2$ annealing.

In any case, the proposed scenario suggests that the defects necessary to stabilize the ferromagnetism are surface related, and thus it is clear that the specific surface of the samples should critically determine their magnetization. This is likely the reason for the common observation of reduced magnetization in bulk materials; on the contrary, nanometric particles may display larger magnetization. Indeed, a large magnetization of ~1.5 $\mu_B$/Mn was reported –up to the best authors knowledge- only for ~5 nm Mn:ZnO nanoparticles [4]. Before concluding, we would like to mention that our analysis has implicitly assumed that the observed ferromagnetism arises from the interaction among magnetic moments of magnetic impurities (Co and Mn in our case), as suggested in Refs. [1,13]. Indeed, we have explored the properties of undoped (Co or Mn free) ZnO particles prepared and processed under the same conditions and treated under $O_2$ or $N_2$ atmospheres, finding in all cases a diamagnetic behaviour without any signature of ferromagnetism. In consequence, the substitution of Zn atoms with magnetic cations appears to be linked to the appearance of ferromagnetism. However, this scenario neglects the possible contribution to the ferromagnetic response of the orbital magnetic moments associated to the perturbation of electronics clouds at the grain surface, a mechanism that was reported to originate weak ferromagnetism in capped-Au nanoparticles [32] and, very recently, in capped ZnO nanoparticles [33]. New experiments should be designed in order to elucidate which of these mechanisms are indeed associated to the appearance of ferromagnetism in these interesting diluted magnetic semiconductors systems.

## 4. Conclusions

In summary, we have shown that both defects and their donor or acceptor character play a major role on the stabilization of ferromagnetism of TM: ZnO oxides. We have proposed that relevant defects –either point-defects or electronic defects associated to gas chemisorption- are likely restricted to a thin nanometric shell covering the particles core, signalling thus a focus for further research on ferromagnetism in ZnO.

**Acknowledgements**

We acknowledge financial support from project MAT2005-5656-C04-01. Quality Chemicals would like to thank CIDEM-Generalitat de Catalunya (contract RDITCRIND04-0153) for financial support. Enlightening discussions with A. Hernando are gratefully acknowledged.

Figure Captions

Figure 1: (a) High-resolution XRD patterns corresponding to ZC1 (upper) and ZM1 (lower) series, fired for 12h in air at temperatures ($T_S$) of 300ºC, 400ºC, 500ºC and 600ºC (the arrows signal the raise of $T_S$). The intensities are shown in log scale. The peaks indicated with (*) reflect the segregation of impurities (($ZnCo)_3O_4$ and $ZnMnO_3$ for ZC1 and ZM1 series, respectively); (b) Concentration of the segregated $(ZnCo)_3O_4$ and $ZnMnO_3$ phases as a function of the firing temperature $T_S$. $(ZnCo)_3O_4$ concentrations (defined as mass of impurity/total mass) were extracted from the corresponding Rietveld refinements, while the amounts of $ZnMnO_3$ were evaluated from the ratio between the 30.2º peak of $ZnMnO_3$ and the 31.7º peak of ZnO.

Figure 2: Room temperature hysteresis loops corresponding to (a) ZM1 and (b) ZC1 samples fired at temperatures between 300ºC and 600ºC. The diamagnetic contribution of the sample holder was carefully discounted in all cases. The insets show the evolution of the magnetization as a function of the temperature for ZM1_500 and ZC1_500 samples (open symbols) and the corresponding fittings (solid lines, see the text for details).

Figure 3: (a) Room temperature magnetization as a function of the applied field corresponding to the aged ZC1_300 sample, annealed in $N_2$ at 300ºC for different times (between 12h and 40h); (b) Room temperature magnetization as a function of the applied field of the aged ZM1_300 sample, annealed in $O_2$ at 300ºC for different times (between 7h and 40h).

Figure 4: Room temperature hysteresis loops corresponding to (a) ZM1_300 sample after successive annealigs in $O_2$ (12h, 300ºC) and $N_2$ (20h, 300ºC), and (b) ZC1_300 sample after successive annealings in $N_2$ (20h, 300ºC) and $O_2$ (20h, 300ºC). It is found that the ferromagnetism in Mn:ZnO can be switched on and off by consecutive annealings in $O_2$ and $N_2$ respectively. The opposite behaviour is found for Co:ZnO. The scheme in (c) summarizes the observed phenomenology.

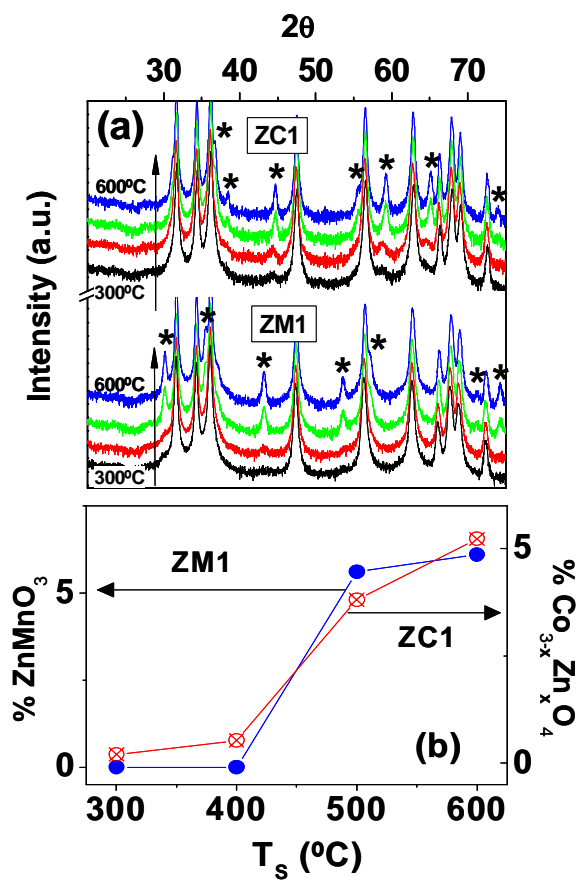

Figure 1

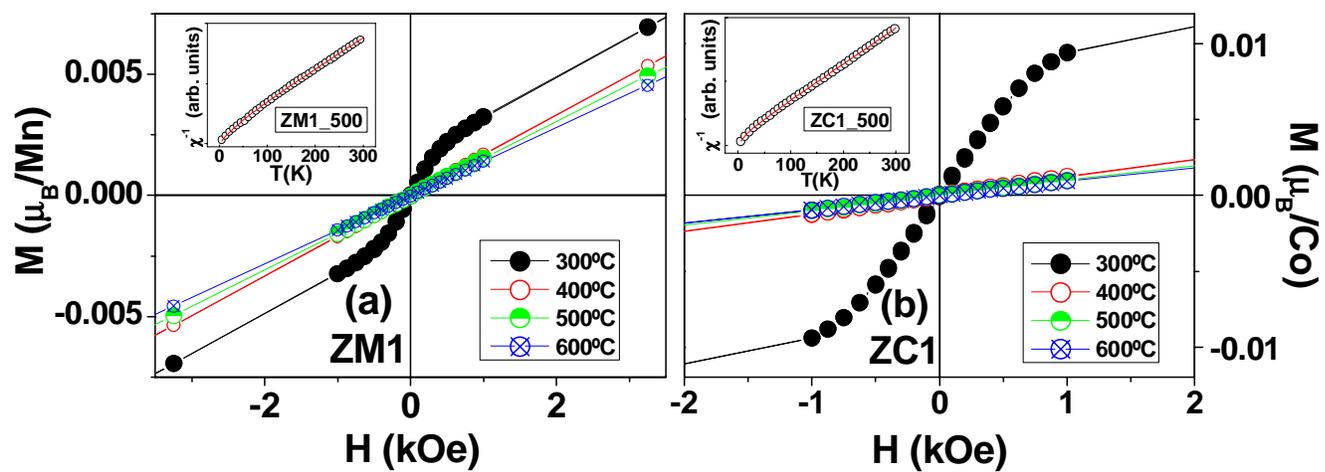

Figure 2

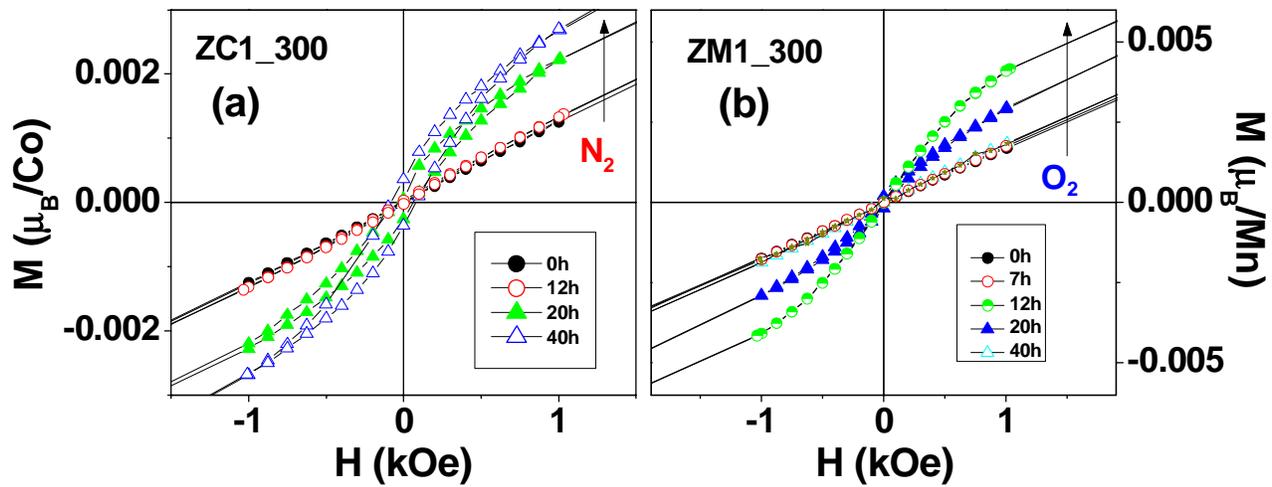

Figure 3

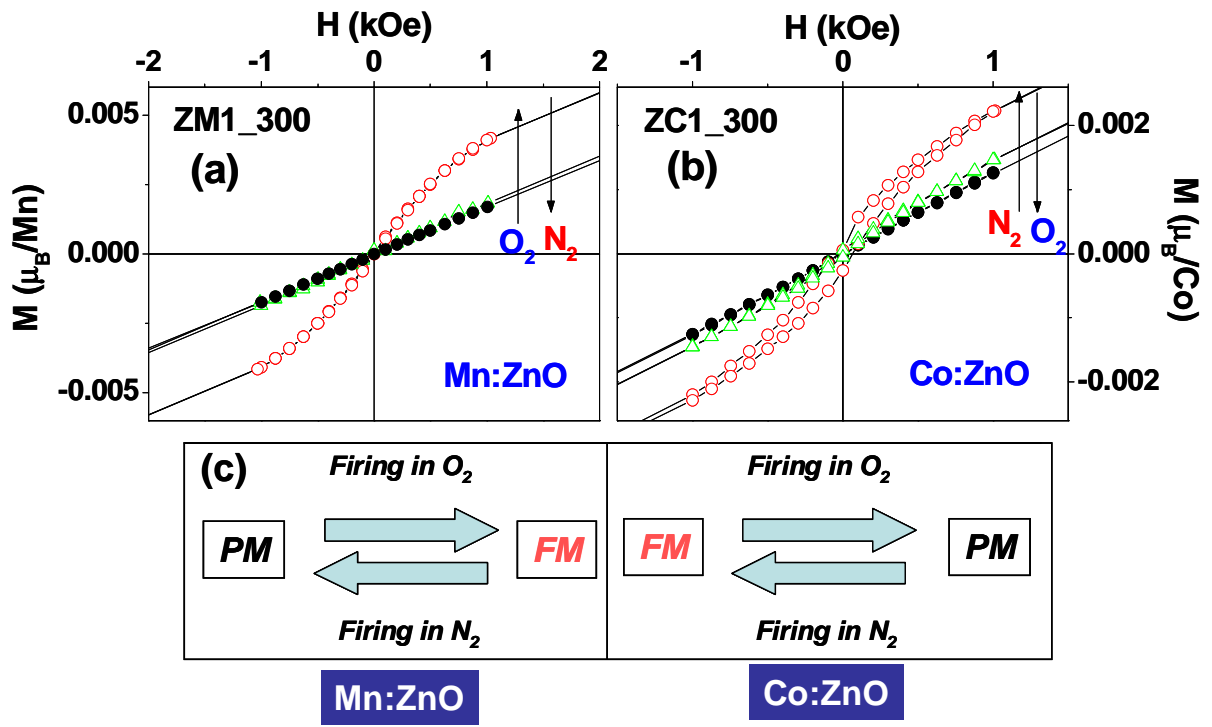

Figure 4